\documentclass[%
 pra,
superscriptaddress,
twocolumn,
showpacs,
 amsmath,amssymb,
 aps,
]{revtex4-1}

\usepackage{graphicx}
\usepackage[space]{grffile}
\usepackage{latexsym}
\usepackage{textcomp}
\usepackage{longtable}
\usepackage{multirow,booktabs}
\usepackage{amsfonts,amsmath,amssymb}
\usepackage{natbib}
\usepackage{url}
\usepackage{hyperref}
\hypersetup{colorlinks=false,pdfborder={0 0 0}}
\usepackage[utf8x]{inputenc}
\usepackage[english]{babel}

\usepackage{graphicx}
\usepackage{dcolumn}
\usepackage{bm}
\usepackage{units}
\usepackage{upgreek}
\usepackage{color}
\usepackage{cancel}
\usepackage{hyperref}
\usepackage{braket}

\newcommand{\ud}{\mathrm{d}}
\newcommand{\ic}{\mathrm{i}}

\newcommand{\zZ}{\mathbb{Z}}

\begin{document}

\title{A quantum random walk of a Bose-Einstein condensate in momentum space}

\author{Gil Summy}
\affiliation{Department of Physics, Oklahoma State University, Stillwater, Oklahoma 74078-3072, USA}

\author{Sandro Wimberger}
\affiliation{DiFeST, Universit\`{a} degli Studi di Parma, Via G. P. Usberti 7/a, 43124 Parma, Italy}
\email{sandro.wimberger@fis.unipr.it}
\affiliation{INFN, Sezione di Milano Bicocca, Gruppo Collegato di Parma, Parma, Italy}
\affiliation{ITP, Heidelberg University, Philosophenweg 12, 69120 Heidelberg, Germany}

\bibliographystyle{apsrev4-1}

\begin{abstract}
Each step in a quantum random walk is typically understood to have two basic components; a `coin-toss' which produces a random superposition of two states, and a displacement which moves each component of the superposition by different amounts. Here we suggest the realization of a walk in momentum space with a spinor Bose-Einstein condensate subject to a quantum ratchet realized with a pulsed, off-resonant optical lattice. By an appropriate choice of the lattice detuning, we show how the atomic momentum can be entangled with the internal spin states of the atoms. For the coin-toss, we propose to use a microwave pulse to mix these internal states. We present experimental results showing an optimized quantum ratchet, and through a series of simulations, demonstrate how our proposal gives extraordinary control of the quantum walk. This should allow for the investigation of possible biases, and classical-to-quantum dynamics in the presence of natural and engineered noise.%
\end{abstract}%


\pacs{03.75.Gg, 05.45.Mt, 05.60.-k, 05.40.Fb}

\maketitle

\section{Introduction} 

Random walks are important in modeling stochastic processes and represent a basic component of diffusion phenomena and non-deterministic motion. Hence it is not surprising that they have broad application in many different contexts of physics and other scientific disciplines \cite{randomwalk}.
The concept of a classical random walk can be translated into a quantum random walk (QRW) \cite{Aharonov1993} using the entanglement between different degrees of freedom. For example, a QRW can be realized by entangling a walk in position space with an intrinsic quantity such as spin \cite{kempe2003}.  In such quantum walks, one degree of freedom typically acts as the `coin' which decides on the direction of the walk. In contrast to its classical counterpart, a quantum coin can produce a superposition of two (or more states) and therefore the corresponding walk is heavily guided by the entanglement between the coin and the walk degree of freedom. A potential application of quantum random walks is probabilistic algorithms for universal quantum computing \cite{Childs2009}.

Based on the pioneering proposal by Aharonov et al. \cite{Aharonov1993}, the authors of \cite{Duer2002} discussed a specific spatial realization of a QRW with cold atoms in optical lattices. A similar setup was realized later by Karski et al. \cite{meschede2009} with single atoms in real space. Our goal is to translate the proposal of \cite{Duer2002} and similar ones (for example \cite{Chan2006,Witthaut}), into a momentum-space random walk, which we will argue has several important advantages. Firstly, the experimental basis of our proposal is the atom-optics kicked rotor (AOKR) which has been studied for more than 20 years and is a well established technique. Secondly, in contrast to other recent work \cite{Preiss2015}, a QRW in momentum space naturally offers the possibility of independently addressing the two-degrees of freedom of the atoms. In the case we discuss, these degrees of freedom would be the internal hyperfine states and the external center-of-mass {\em momenta} of atoms in a Bose-Einstein condensate (BEC).

The realization of a coin-operator is relatively straightforward and for our system could be implemented with resonant micro-wave radiation. The major difficulty of a QRW lies in producing a shift in momentum space that is dependent on an atom's internal state. That is, we need a shift operator that takes the form 
$$\hat T = \exp(i \hat x \Delta p/\hbar) |1\rangle \langle 1|+\exp(-i \hat x \Delta p/\hbar) |2\rangle \langle 2|\,,$$ 
which shifts the momentum by $\pm\Delta p$ depending on whether the atom resides in the internal state $|1\rangle$ or $|2\rangle$. In the periodic potential of an optical lattice, momentum is naturally quantized in units of two atomic recoils $2p_{\rm R}=2\hbar k_L$, $k_L$ being the wave vector of the laser creating the lattice. Expressing momentum in these units, the shift operator becomes 
$$\hat T = \exp(i \hat \theta n) |1\rangle \langle 1|+\exp(-i \hat \theta n) |2\rangle \langle 2|\,,$$ 
where $n$ is integer. In the usual random walk setup, $n=1$ which corresponds to nearest neighbor coupling in momentum space. We propose to realize the shift by kicking a BEC with a periodic lattice. Such systems are routinely realized in the context of the AOKR \cite{RaizenAdv,SW2011}, a standard model for investigating quantum chaos and Anderson localization in momentum space \cite{Izr1990}. The kick will indeed act as a biased shift, which depends on the internal state of the atom, when {\em (i)} employing quantum resonance conditions on the dynamics \cite{Izr1990, SW2011} and {\em (ii)} destroying the spatial-temporal symmetry using a quantum ratchet \cite{sadgrove2007, gil2008, SW2011, gil2012}. The direction of the walk is then controlled by the sign of the kick potential, which itself is controlled by the internal state. We now explain in detail how to implement a QRW in momentum space along these lines.

\section{Quantum walks at quantum resonance}  
\label{sec:2}

\begin{figure}[tb!]
\begin{center}
\includegraphics[width=\columnwidth]{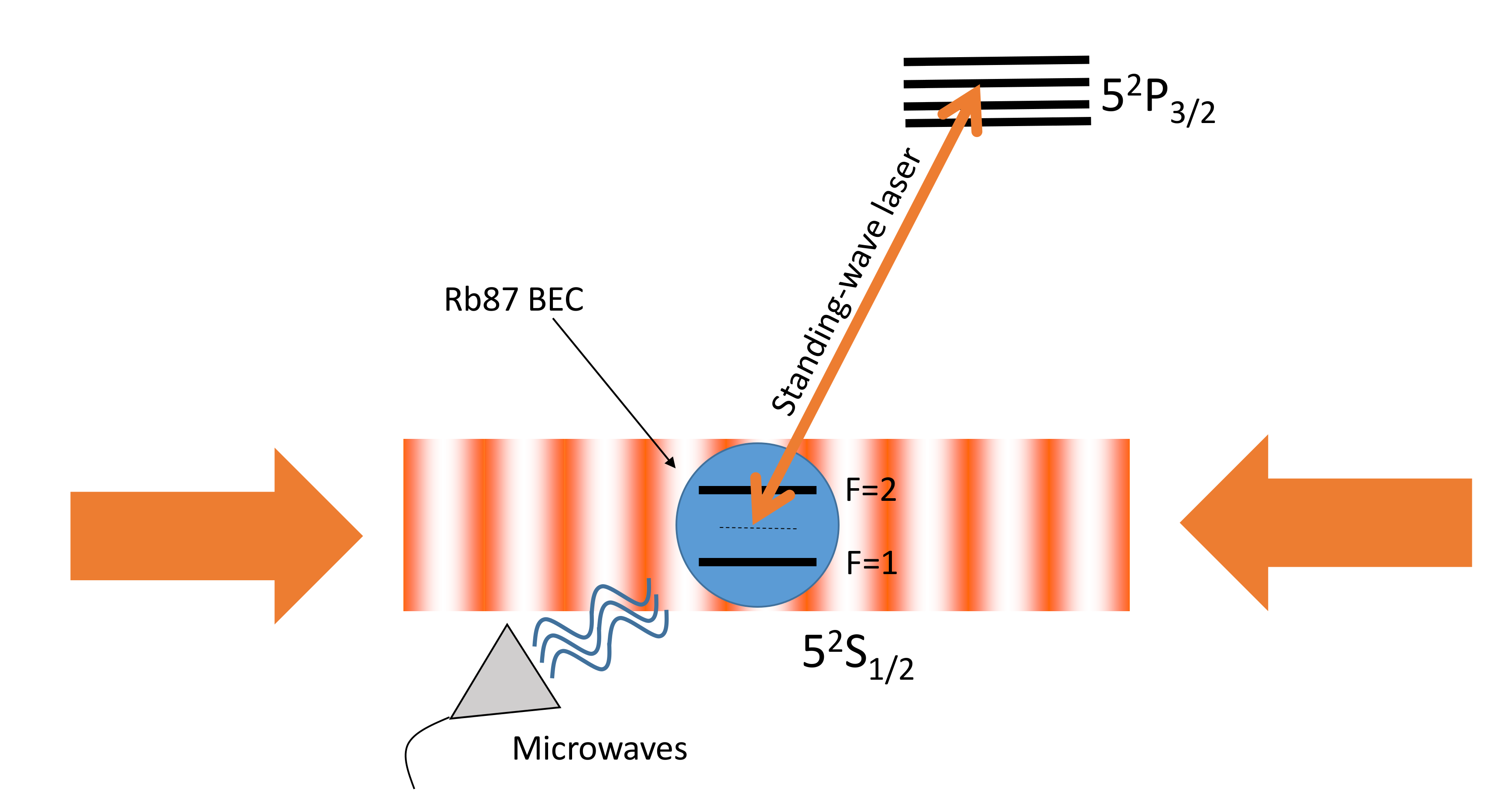}
\caption{\label{fig:0} (color online). Schematic of our proposed experiment for the realization of a quantum random walk in momentum space. The optical lattice is pulsed periodically to implement the momentum shifts at quantum resonance. The internal states $F=1$ and $F=2$ of the atoms in the rubidium 87 condensate are controlled by microwaves.%
}
\end{center}
\end{figure}

AOKR experiments work with ultracold atoms subject to periodic kicks by an optical lattice. A schematic of the experimental setup we have used in the past is shown in Fig. \ref{fig:0}. The small initial temperature necessary to resolve the single momentum peaks is most easily reached using a Bose-Einstein condensate. For sufficiently dilute condensates we may safely neglect atom-atom interactions. Using dimensionless variables, the quantum dynamics of the center-of-mass of the atoms are then described by the following single-particle Hamiltonian \cite{RaizenAdv,SW2011}:
\begin{equation}
  \label{eq:Hkick}
  \hat{H}(\hat{x},\hat{p}_x,t)\;=\;\frac{\hat{p}_x^2}{2} + k\cos( \hat{x}) \sum_{j\in \zZ} \delta(t-j\tau)\ .
\end{equation}
Here $j$ counts the number of kicks, the kick period is $\tau$, the kick strength is $k=\Omega^2 \tau_{\rm p}/\Delta$, where $\tau_p \ll \tau$ is the pulse length, $\Omega$ is the Rabi frequency, and $\Delta$ is the detuning of the kicking laser from the atomic transition.

The periodicity of the potential implies conservation of quasimomentum (QM) $\beta$ with $p_x = n + \beta$, where $n$ is integer in our units and $\beta$ takes on values between $0$ and $1$. Using Bloch theory, the atom dynamics from immediately before the $(j-1)$-th kick to immediately before the next $j$-th kick is given by the Floquet operator \cite{SW2011}:
\begin{equation}
  \hat{\cal U}_{\beta,k} \;=\; e^{-\ic \tau(\hat{\cal N}+ \beta)^2/2} \; e^{-\ic k \cos(\hat{\theta})} \ ,
 \label{eq:Fl}
\end{equation}
where $\hat{\cal N}=- \ic \frac{\ud}{\ud\theta}$ is the (angular) momentum operator with periodic boundary conditions and $\theta=x \  {\rm mod}(2\pi)$. The second factor of the Floquet operator can be expressed in momentum representation as 
\begin{equation}
\exp\left(-\ic k \cos\theta \right) = \sum\limits_{m}(-\ic)^{m}\exp(- \ic m\theta){J_{m}(k)}
 \label{eq:bessel}
\end{equation}
where the $J$'s are Bessel functions of the first kind and give the coupling amplitudes between the initial $n$ and final $m$ momentum states. The Bessel function properties are such that this amplitude will decay rapidly as the difference $|m-n|$ increases \cite{Izr1990}. In fact for $k \sim 1$ roughly only nearest neighbor momentum states are coupled such that $m= n \pm 1$. However, the symmetric nature of the momentum step and the fact that there is no role for a coin-toss in this setup makes it difficult to implement a QRW walk in its usual form.

In the following, we require quantum resonant dynamics of the AOKR. This implies that the first factor on the right of Eq. \eqref{eq:Fl} equals the identity. The principal quantum resonances are obtained for $\tau=2\pi\ell$, with positive integer $\ell$, and $\beta=1/2 + l/\ell$, with $l=0,1,\ldots,\ell-1$ \cite{Izr1990,nonl,SW2011}. The quantum resonances can be seen as the Talbot effect (albeit in the time domain) for atomic matter waves diffracted from the optical grating induced by the flashed periodic potential \cite{PhysRevLett.83.5407,SW2011}. Examples are $\tau=2\pi$ (for $\beta=0.5$) or $\tau=4\pi$ (for $\beta=0$), corresponding to the half or the full Talbot time, respectively. For realizing a perfect atomic ratchet, quantum resonance conditions should be met, see \cite{gil2008, SW2011, SadgroveWimberger2009, gil2012}. Quantum walks based on the Talbot effect \cite{BK2004} and quantum accelerator modes \cite{QAM2006} were proposed, yet never realized due to technical problems in their implementation. We will now describe in detail how to implement a simpler QRW at quantum resonance using an atomic ratchet current whose direction is controlled by two different internal states of the atoms.

\begin{figure}[tb!]
\begin{center}
\includegraphics[width=\columnwidth]{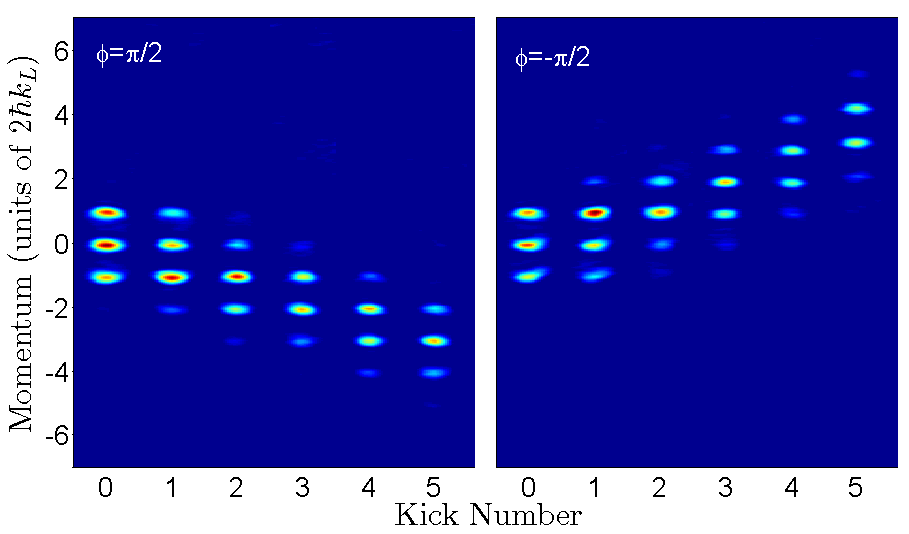}
\caption{\label{fig:00} (color online). Experimental data showing the momentum distribution of a BEC as a function of time for an asymmetric initial state of the form $|\psi_3 \rangle$. The ratchet effect and its dependence on the relative phase $\phi$ are very clear. Note that in contrast to the experiments in Ref. \cite{gil2008}, almost all of the initial state participates in the ratchet. The data are taken in the limit of small kick periods simulating kicks at (half) the Talbot time while keeping decoherence effects to the minimum, see \cite{Sadgrove_2005}.%
}
\end{center}
\end{figure}

The dynamics given by Eq. \eqref{eq:Hkick} can be made asymmetric in $n$ space by breaking the spatial-temporal symmetry of the problem. Experimentally, this is most easily realized by the choice of the initial state, such as 
$$\ket{\psi_2}=\ket{\psi(n,t=0)}=\frac{1}{\sqrt{2}}(\ket{n=0}+e^{i \phi}\ket{n=1})\,.$$ 
Such a state receives an average change in momentum per kick of 
\begin{equation}
\Delta \langle  \hat p \rangle=-k\sin( \phi)/2\\,
 \label{eq:pmean}
\end{equation}
 so that by choosing $\phi=\pm \pi/2$ and $k \sim 2$ the average momentum can be either decreased (increased) by one step \cite{gil2008, SW2011, SadgroveWimberger2009, gil2012}. New experimental data for an initial state with three components, i.e., 
 $$\ket{\psi_3} =\frac{1}{\sqrt{3}}(e^{-i \phi}\ket{-1}+\ket{0}+e^{i \phi}\ket{1})\,,$$ 
 are shown in Fig. \ref{fig:00}. The directed transport is clearly visible, as well as its directional dependence on the phase $\phi$. Including more momentum states improves the `purity' of the ratchet effect. For example, in \cite{gil2008} a significant amount of the initial two component state recoiled in the opposite direction of the ratchet. This can be contrasted with the almost pure ratchet demonstrated by the data in Fig. \ref{fig:00}. Below we will also consider the initial state 
 $$\ket{\psi_4}=\frac{1}{2}(e^{i \frac{\pi}{2}}\ket{-1}+\ket{0}+e^{-i \frac{\pi}{2}}\ket{1}+e^{-i \pi}\ket{2})\,.$$

As mentioned previously, we want to make a step for our walk contingent on the result of a coin-toss. Here we propose to use a coin-toss that connects the components of a  pseudo-spin $\pm 1/2$ system that experimentally corresponds to the ground hyperfine levels of a rubidium 87 atom. Such an operation can be implemented in the lab with a microwave pulse resonant to the transition between the $F=1, m_F=0$ and $F=2, m_F=0$ levels of the $5^{2}S_{1/2}$ state, see Fig. \ref{fig:0}. A 50-50 coin toss in such a scheme would correspond to a $\pi/2$ pulse of the microwaves.

We now want to engineer our system such that the hyperfine level controls the direction of the kick. That is, the one-step operator should be expressible as
\begin{equation}
\hat U_{\rm kick} = \exp\left(- \ic k \cos( {\hat \theta} ) \sigma_z \right) \,,
\label{eq:sigma1}
\end{equation}
where $\sigma_z$ is the Pauli matrix. 

We note that in our previous experiments with kicked rubidium 87 BECs \cite{Behinaein_2006,gil2008,Talukdar_2010,gil2012,gil2013a}, the BECs were prepared in the $5^{2}S_{1/2} F=1$ level and the kicking light had a frequency corresponding to transitions between the $5^{2}S_{1/2} F=2$ and $5^{2}P_{3/2} F=3$ levels. This produced a standing wave with a detuning of $\Delta\sim 6.8$ GHz. Clearly this configuration can no longer be applied to what we wish to achieve here as the light would be resonant with one of the internal states of interest.  However, by detuning the kick laser frequency between the two hyperfine levels, we can regain the far off-resonant condition and produce periodic potentials. Then the ac-Stark shift, and hence also the kick strength $k$, see its definition after Eq. \eqref{eq:Hkick}, differ in sign for the two states since the laser is either red or blue detuned ($\Delta>0$ or $\Delta<0$). What this implies for our proposal is that the ratchet current changes sign with the sign change in the detuning. For example, when the $F=1$ component has $k>0$ and a negative ratchet current, the $F=2$ level will experience $k<0$ and a positive ratchet current, see Eq. \eqref{eq:pmean}.

The internal degree of freedom is denoted by spin up, $\ket{1/2}$, and spin down, $\ket{-1/2}$ (experimentally corresponding to the $F=2, m_F=0$ and $F=1, m_F=0$ states), while the interaction between the spins is represented by the two-parameter unitary rotation matrix
\begin{equation}
 {\bf M }(\alpha, \chi) = \frac{1}{\sqrt{2}} \begin{pmatrix}
\cos{\frac{\alpha}{2}} &  e^{-\ic \chi}\sin{\frac{\alpha}{2}} \\ -e^{\ic \chi}\sin{\frac{\alpha}{2}} & \cos{\frac{\alpha}{2}}  \end{pmatrix} \,.
 \label{eq:uni}
\end{equation}
Before the kicking sequence, we propose to initialize the system starting from the spin down state and an application of a so-called Hadamard gate (the matrix above with $\alpha=\pi/2$ and $\chi=0$). Hence, we start the first step with the internal state
\begin{equation}
 {\bf \hat M}(\pi/2,0) \ket{n,s=-1/2} =  \frac{1}{\sqrt{2}}\left( \ket{n,-1/2} + \ket{n,1/2} \right)\,.
 \label{eq:init}
\end{equation}
The matrix for the single coin toss applied after each kick (or step of the walk) is most conveniently represented by the matrix of a 50-50 beam splitter acting on $\ket{n,s}$, e.g. by
$$ {\bf M}(\pi/2,-\pi/2) = \frac{1}{\sqrt{2}} \begin{pmatrix} 1 & \ic \\ \ic & 1 \end{pmatrix}\,.$$ 
 Note that this choice has the advantage of being symmetric with respect to the internal initial state \cite{kempe2003}. After each kick, ${\bf M}(\pi/2,-\pi/2)$ acts on the internal state, which produces a strong mixing of internal and external degrees of freedom during the temporal evolution.

\section{Numerical results} 

\begin{figure}[tb!]
\begin{center}
\includegraphics[width=\columnwidth]{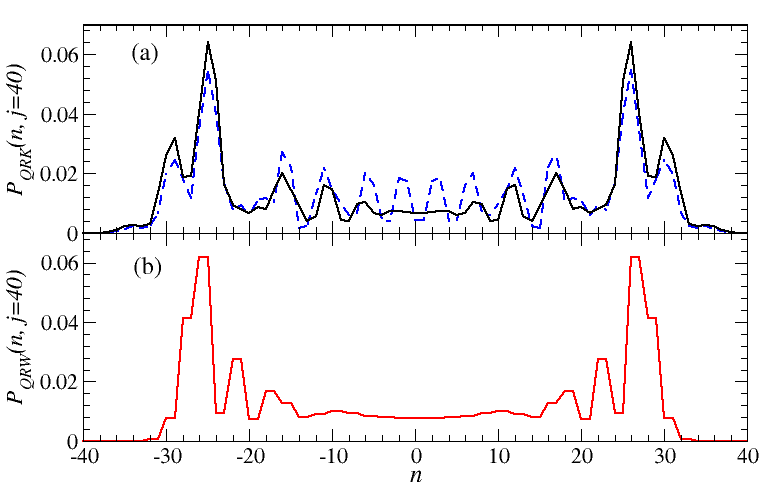}
\caption{\label{fig:3} (color online). Numerical simulations comparing the AOKR walk, see panel (a), to a standard QRW in momentum space with shift operator $\hat T_1$ from Eq. \eqref{eq:shiftT}, see panel (b), both after $40$ steps. In (a) $k=1.5$ for two different initial states of the AOKR in its external degree of freedom $\ket{\psi_2}$ with $\phi=-\pi/2$ (dashed line) and $\ket{\psi_4}$ (solid line).%
}
\end{center}
\end{figure}

Our experimental observables are the internal-state resolved momentum distributions $P_s(n)$ of the atoms. Thus for an arbitrary state of the full system 
$$\ket{\psi(j)}= \sum_{n,s} c_{n,s}(j) \ket{n,s}\,,$$ 
we can measure 
\begin{equation}
P_{-1/2}(n,j) = |c_{n,s=-1/2}|^2 
\label{eq:dist1}
\end{equation}
and 
\begin{equation}
P_{1/2}(n,j) = |c_{n,s=1/2}|^2\,.
\label{eq:dist2}
\end{equation}
Figure \ref{fig:3} (a) shows the total momentum distributions 
\begin{equation}
P(n,j) = P_{-1/2}(n,j) + P_{1/2}(n,j)
\label{eq:dist3}
\end{equation}
obtained for our AOKR realization for two different types of initial motional states,  while Fig. \ref{fig:3}(b) presents the standard QRW with shift operator 
\begin{equation}
\hat T_1 = \exp(i \hat \theta) |1\rangle \langle 1|+\exp(-i \hat \theta) |2\rangle \langle 2|\,.
\label{eq:shiftT}
\end{equation}

Overall our proposed realization of a walk has the same features as the standard QRW, with strong peaks at the maxima which move ballistically outward such that $|n_{\rm max}| \propto j$. To make the comparison more meaningful, our new method requires the insertion of a prefactor in the previous relation because the Bessel functions cause a coupling between states other than just nearest neighbor [see discussion around Eq. \eqref{eq:bessel}]. The overall coupling strength that best matched the standard QRW in Fig. \ref{fig:3}(b) was $k \approx 1.5$. 

We also draw attention to the fact that the oscillations around the center of the distributions can be suppressed by choosing an initial state composed of more momentum states [see solid line as compared to the dashed line in Fig. \ref{fig:3}(a)]. Interestingly, the final result is very stable with respect to the phase choice $\phi$ in the initial state, which can be detuned by up to $10\ldots20$ percent without noticeable differences for our observation times.

\section{Quantum-to-classical transition of the walks} 

\begin{figure}[tb!]
\begin{center}
\includegraphics[width=\columnwidth]{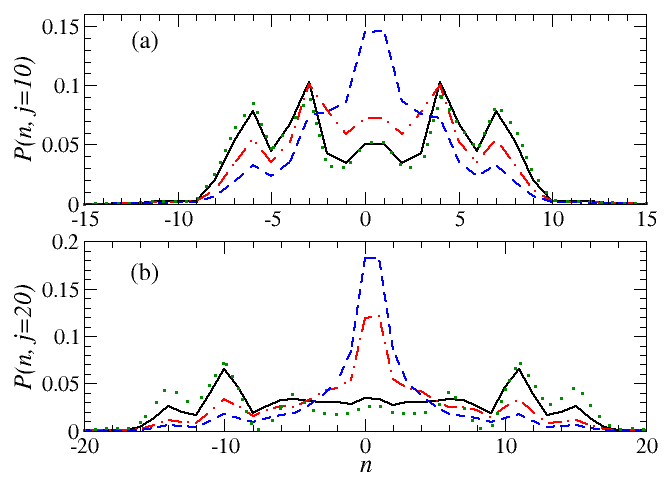}
\caption{\label{fig:4} (color online). Numerical simulations on the impact of a finite width in QM on a QRW in momentum space for the initial state $\ket{\psi_2}$ after (a) $10$ and (b) $20$ steps. The ideal quantum walk with resonant QM is shown by the green dotted lines. Walks with a finite QM distribution (incoherently averaged over $10^4$ values corresponding to a typical atom number in the BEC) are shown for $\Delta \beta=0.01$ (black solid line), $0.025$ (red dot-dashed line) and $0.05$ (blue dashed line).%
}
\end{center}
\end{figure}

Our QRW becomes classical (manifested by the appearance of a Gaussian limit distribution around zero momentum) when adding dephasing. Randomizing the mixing between the two internal states during the coin toss leads to such a result, with the characteristic standard deviation for a classical walk growing as $\sqrt{j}$, as we checked (not shown here).  Another more natural source of dephasing for our kind of experiment arises from deviations in QM from the resonant value, see the discussion in section \ref{sec:2}. Any real BEC has some finite width in (quasi)momentum, which is typically about $\Delta \beta = 0.01$ in its full width at half maximum (FWHM) \cite{Ryu}. The dependence of the walk on a finite width in QM is shown in Fig. \ref{fig:4} for different kick numbers, but otherwise the same parameters as used in Fig. \ref{fig:3}. Up to about 10 to 20 kicks, typical widths of $0.01$ have little effect on the quantum walk, whereas larger widths induce a transition to a classical walk in a systematic fashion. 

We conclude that a QRW could indeed be realized with a sufficiently small initial width in QM, which is guaranteed by modern setups with Bose-Einstein condensates. On the other hand, by actively controlling the width in QM, the sensitive dependence of the walk may in turn be used as a reliable detector of decoherence. Consequently, our proposal can be readily extended to investigate fundamental quantum decoherence processes and their impact on QRWs. The sensitive dependence of the AOKR dynamics on QM was also exploited, e.g., in \cite{gil2013a} to determine the initial momentum width of a condensate.

\begin{figure}[tb!]
\begin{center}
\includegraphics[width=\columnwidth]{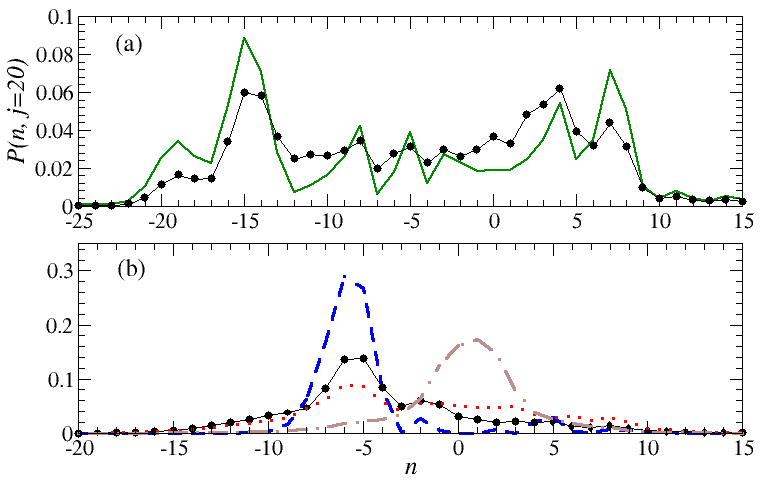}
\caption{\label{fig:5} (color online). Numerical results on a biased QRW with $k_{-1/2}=-1.72$ ($F=1$) and $k_{1/2}=1$ ($F=2$). Data are shown by the solid lines after $j=20$ steps for the initial state $\ket{\psi_4}$. The black filled circles in (a) present corresponding results for a distribution of QM with FWHM\,$=0.01$. The dashed line in (b) corresponds to an optimally directed walk obtained when applying a $\pi$ pulse at half of the evolution time, i.e., at the step $j=10$. The other data sets in (b) show results for the same protocol but with finite QM distributions of $\Delta \beta=0.01$ (black filled circles), $0.025$ (red dotted line) and $0.25$ (brown dot-dashed line).%
}
\end{center}
\end{figure}

\section{Realization of biased walks} 

Our setup permits us to investigate a biased quantum walk in momentum space. Such a walk is realized by choosing a laser wavelength for the effective kick potentials which has two different (but again oppositely signed) detunings from the excited level. Since the kick strengths are inversely proportional to the detunings, their ratio $k_{-1/2}/k_{1/2}$ is given by the inverse ratio of the detunings. This generalizes the one-step operator from Eq. \eqref{eq:sigma1} into
\begin{equation}
\label{eq:sigma2}
\hat U_{\rm kick} = \exp \left( -\ic \cos(\hat{\theta})
\begin{pmatrix}
k_{1/2} &  0 \\  0 &   k_{-1/2}
\end{pmatrix} \right) \,,
\end{equation}
where the bias is controlled at will by the ratio $k_{-1/2}/k_{1/2}$.

We present numerical data of such a biased QRW in Fig. \ref{fig:5}, which also contains results for finite distributions of QM. Interestingly, the quantum walk can be steered into one direction by applying an additional $\pi$ pulse [using ${\bf M}(\pi,0)$] to the internal degree of freedom after exactly half the steps, see the dashed line in Fig. \ref{fig:5}(b). The speed of such a walk is controlled by the difference of kick strengths between the two internal states, i.e. the larger the difference the faster the peak moves toward the left in our case. Increasing the FWHM of the QM distribution turns the quantum walk classical again, as was seen in Fig. \ref{fig:4}. This is visible in Fig. \ref{fig:5}(b), where the peak that has been at finite momentum for resonant QM moves toward zero momentum as would be the case for a classical unbiased diffusive walk. The symmetry of the walk can partly be restored by applying the $\pi$ pulse after a different fraction of the kicks, i.e., after $5$ steps rather than $10$ as in Fig. \ref{fig:5}(b). However the total spreading is then slower than the situation represented by Fig. \ref{fig:5}(a) (without the population inversion between the kick sequence). 

The quantum walk obtained by inversion is very stable at $\chi=0$ with respect to the precise value of $\alpha$ defined in Eq. \eqref{eq:uni}, which just affects the height of the leftward moving peak, rather than changing the overall momentum distribution. These possibilities for controlling the walk, together with its robustness against parameter variation, make it interesting for actual implementations and use in quantum information applications.

\section{Conclusions} 
\label{sec:concl}

The proposed realization of a QRW in momentum space has several advantages with respect to previous implementation of a quantum walk. With current setups, which allow for a detection window of about 50 momentum states \cite{Behinaein_2006,gil2008,gil2012,gil2013b}, walks between 10 and 50 steps could be experimentally implemented. 

We can easily tune the relative weights in the walk in order to bias it, simply by changing the relative detunings from the hyperfine levels. Moreover, the single particle walks studied in \cite{meschede2009,PhysRevLett.103.090504,PhysRevLett.104.153602,PhysRevLett.104.100503,PhysRevLett.104.153602},  are not easily extended to a many-body setup \cite{PhysRevLett.108.010502}. However in contrast to \cite{Preiss2015}(which addresses two-body correlations but not the internal states of the atoms), our implementation works with fully controllable access to both external and internal degrees of freedom. 

Quantum walks in momentum space are also useful for investigating decoherence during the walk and the quantum to classical transition, see e.g. \cite{njpBonn} and references therein, in general by adding noise to one or both degrees of freedom in a controlled manner. Apart from the engineering of quantum transport, possible future applications of our proposed walks in momentum space are the test of fundamental quantum relations in measurements theory \cite{njpBonn} and entanglement statistics \cite{Ra22012013,bosonsample2,bosonsample5}.

\begin{acknowledgments}
We thank Jiating Ni and Wakun Lam for assistance obtaining the experimental results and Mark Sadgrove and Shmuel Fishman for useful comments on the manuscript. SW acknowledges financial support by the FIL program of Parma University.
\end{acknowledgments}

\bibliographystyle{apsrev4-1}

%


\begin{thebibliography}{29}%
\makeatletter
\providecommand \@ifxundefined [1]{%
 \@ifx{#1\undefined}
}%
\providecommand \@ifnum [1]{%
 \ifnum #1\expandafter \@firstoftwo
 \else \expandafter \@secondoftwo
 \fi
}%
\providecommand \@ifx [1]{%
 \ifx #1\expandafter \@firstoftwo
 \else \expandafter \@secondoftwo
 \fi
}%
\providecommand \natexlab [1]{#1}%
\providecommand \enquote  [1]{``#1''}%
\providecommand \bibnamefont  [1]{#1}%
\providecommand \bibfnamefont [1]{#1}%
\providecommand \citenamefont [1]{#1}%
\providecommand \href@noop [0]{\@secondoftwo}%
\providecommand \href [0]{\begingroup \@sanitize@url \@href}%
\providecommand \@href[1]{\@@startlink{#1}\@@href}%
\providecommand \@@href[1]{\endgroup#1\@@endlink}%
\providecommand \@sanitize@url [0]{\catcode `\\12\catcode `\$12\catcode
  `\&12\catcode `\#12\catcode `\^12\catcode `\_12\catcode `\%12\relax}%
\providecommand \@@startlink[1]{}%
\providecommand \@@endlink[0]{}%
\providecommand \url  [0]{\begingroup\@sanitize@url \@url }%
\providecommand \@url [1]{\endgroup\@href {#1}{\urlprefix }}%
\providecommand \urlprefix  [0]{URL }%
\providecommand \Eprint [0]{\href }%
\providecommand \doibase [0]{http://dx.doi.org/}%
\providecommand \selectlanguage [0]{\@gobble}%
\providecommand \bibinfo  [0]{\@secondoftwo}%
\providecommand \bibfield  [0]{\@secondoftwo}%
\providecommand \translation [1]{[#1]}%
\providecommand \BibitemOpen [0]{}%
\providecommand \bibitemStop [0]{}%
\providecommand \bibitemNoStop [0]{.\EOS\space}%
\providecommand \EOS [0]{\spacefactor3000\relax}%
\providecommand \BibitemShut  [1]{\csname bibitem#1\endcsname}%
\let\auto@bib@innerbib\@empty

\bibitem [{\citenamefont {Weiss}(1994)}]{randomwalk}%
  \BibitemOpen
  \bibfield  {author} {\bibinfo {author} {\bibfnamefont {G.~H.}\ \bibnamefont
  {Weiss}},\ }\href@noop {} {\emph {\bibinfo {title} {{Aspects and Applications
  of the Random Walk}}}}\ (\bibinfo  {publisher} {North Holland},\ \bibinfo
  {address} {Amsterdam},\ \bibinfo {year} {1994})\BibitemShut {NoStop}%
\bibitem [{\citenamefont {Aharonov}\ \emph {et~al.}(1993)\citenamefont
  {Aharonov}, \citenamefont {Davidovich},\ and\ \citenamefont
  {Zagury}}]{Aharonov1993}%
  \BibitemOpen
  \bibfield  {author} {\bibinfo {author} {\bibfnamefont {Y.}~\bibnamefont
  {Aharonov}}, \bibinfo {author} {\bibfnamefont {L.}~\bibnamefont
  {Davidovich}}, \ and\ \bibinfo {author} {\bibfnamefont {N.}~\bibnamefont
  {Zagury}},\ }\href {\doibase 10.1103/PhysRevA.48.1687} {\bibfield  {journal}
  {\bibinfo  {journal} {Phys. Rev. A}\ }\textbf {\bibinfo {volume} {48}},\
  \bibinfo {pages} {1687} (\bibinfo {year} {1993})}\BibitemShut {NoStop}%
\bibitem [{\citenamefont {Kempe}(2003)}]{kempe2003}%
  \BibitemOpen
  \bibfield  {author} {\bibinfo {author} {\bibfnamefont {J.}~\bibnamefont
  {Kempe}},\ }\href@noop {} {\bibfield  {journal} {\bibinfo  {journal}
  {Contemporary Physics}\ }\textbf {\bibinfo {volume} {44}},\ \bibinfo {pages}
  {307} (\bibinfo {year} {2003})}\BibitemShut {NoStop}%
\bibitem [{\citenamefont {Childs}(2009)}]{Childs2009}%
  \BibitemOpen
  \bibfield  {author} {\bibinfo {author} {\bibfnamefont {A.~M.}\ \bibnamefont
  {Childs}},\ }\href {\doibase 10.1103/PhysRevLett.102.180501} {\bibfield
  {journal} {\bibinfo  {journal} {Phys. Rev. Lett.}\ }\textbf {\bibinfo
  {volume} {102}},\ \bibinfo {pages} {180501} (\bibinfo {year}
  {2009})}\BibitemShut {NoStop}%
\bibitem [{\citenamefont {D\"ur}\ \emph {et~al.}(2002)\citenamefont {D\"ur},
  \citenamefont {Raussendorf}, \citenamefont {Kendon},\ and\ \citenamefont
  {Briegel}}]{Duer2002}%
  \BibitemOpen
  \bibfield  {author} {\bibinfo {author} {\bibfnamefont {W.}~\bibnamefont
  {D\"ur}}, \bibinfo {author} {\bibfnamefont {R.}~\bibnamefont {Raussendorf}},
  \bibinfo {author} {\bibfnamefont {V.~M.}\ \bibnamefont {Kendon}}, \ and\
  \bibinfo {author} {\bibfnamefont {H.-J.}\ \bibnamefont {Briegel}},\ }\href
  {\doibase 10.1103/PhysRevA.66.052319} {\bibfield  {journal} {\bibinfo
  {journal} {Phys. Rev. A}\ }\textbf {\bibinfo {volume} {66}},\ \bibinfo
  {pages} {052319} (\bibinfo {year} {2002})}\BibitemShut {NoStop}%
\bibitem [{\citenamefont {Karski}\ \emph {et~al.}(2009)\citenamefont {Karski},
  \citenamefont {F\"orster}, \citenamefont {Choi}, \citenamefont {Steffen},
  \citenamefont {Alt}, \citenamefont {Meschede},\ and\ \citenamefont
  {Widera}}]{meschede2009}%
  \BibitemOpen
  \bibfield  {author} {\bibinfo {author} {\bibfnamefont {M.}~\bibnamefont
  {Karski}}, \bibinfo {author} {\bibfnamefont {L.}~\bibnamefont {F\"orster}},
  \bibinfo {author} {\bibfnamefont {J.-M.}\ \bibnamefont {Choi}}, \bibinfo
  {author} {\bibfnamefont {A.}~\bibnamefont {Steffen}}, \bibinfo {author}
  {\bibfnamefont {W.}~\bibnamefont {Alt}}, \bibinfo {author} {\bibfnamefont
  {D.}~\bibnamefont {Meschede}}, \ and\ \bibinfo {author} {\bibfnamefont
  {A.}~\bibnamefont {Widera}},\ }\href@noop {} {\bibfield  {journal} {\bibinfo
  {journal} {Science}\ }\textbf {\bibinfo {volume} {325}},\ \bibinfo {pages}
  {174} (\bibinfo {year} {2009})}\BibitemShut {NoStop}%
\bibitem [{\citenamefont {Chandrashekar}(2006)}]{Chan2006}%
  \BibitemOpen
  \bibfield  {author} {\bibinfo {author} {\bibfnamefont {C.~M.}\ \bibnamefont
  {Chandrashekar}},\ }\href {\doibase 10.1103/PhysRevA.74.032307} {\bibfield
  {journal} {\bibinfo  {journal} {Phys. Rev. A}\ }\textbf {\bibinfo {volume}
  {74}},\ \bibinfo {pages} {032307} (\bibinfo {year} {2006})}\BibitemShut
  {NoStop}%
  \bibitem{Witthaut} D. Witthaut, Phys. Rev. A {\bf 82}, 033602 (2010).
  
\bibitem [{\citenamefont {Preiss}\ \emph {et~al.}(2015)\citenamefont {Preiss},
  \citenamefont {Ma}, \citenamefont {Tai}, \citenamefont {Lukin}, \citenamefont
  {Rispoli}, \citenamefont {Zupancic}, \citenamefont {Lahini}, \citenamefont
  {Islam},\ and\ \citenamefont {Greiner}}]{Preiss2015}%
  \BibitemOpen
  \bibfield  {author} {\bibinfo {author} {\bibfnamefont {P.~M.}\ \bibnamefont
  {Preiss}}, \bibinfo {author} {\bibfnamefont {R.}~\bibnamefont {Ma}}, \bibinfo
  {author} {\bibfnamefont {M.~E.}\ \bibnamefont {Tai}}, \bibinfo {author}
  {\bibfnamefont {A.}~\bibnamefont {Lukin}}, \bibinfo {author} {\bibfnamefont
  {M.}~\bibnamefont {Rispoli}}, \bibinfo {author} {\bibfnamefont
  {P.}~\bibnamefont {Zupancic}}, \bibinfo {author} {\bibfnamefont
  {Y.}~\bibnamefont {Lahini}}, \bibinfo {author} {\bibfnamefont
  {R.}~\bibnamefont {Islam}}, \ and\ \bibinfo {author} {\bibfnamefont
  {M.}~\bibnamefont {Greiner}},\ }\href {\doibase 10.1126/science.1260364}
  {\bibfield  {journal} {\bibinfo  {journal} {Science}\ }\textbf {\bibinfo
  {volume} {347}},\ \bibinfo {pages} {1229} (\bibinfo {year}
  {2015})}\BibitemShut {NoStop}%
\bibitem [{\citenamefont {Raizen}(1999)}]{RaizenAdv}%
  \BibitemOpen
  \bibfield  {author} {\bibinfo {author} {\bibfnamefont {M.~G.}\ \bibnamefont
  {Raizen}},\ }\href@noop {} {\bibfield  {journal} {\bibinfo  {journal} {Adv.
  At. Mol. Opt. Phys.}\ }\textbf {\bibinfo {volume} {41}},\ \bibinfo {pages}
  {43} (\bibinfo {year} {1999})}\BibitemShut {NoStop}%
\bibitem [{\citenamefont {Sadgrove}\ and\ \citenamefont
  {Wimberger}(2011)}]{SW2011}%
  \BibitemOpen
  \bibfield  {author} {\bibinfo {author} {\bibfnamefont {M.}~\bibnamefont
  {Sadgrove}}\ and\ \bibinfo {author} {\bibfnamefont {S.}~\bibnamefont
  {Wimberger}},\ }\href@noop {} {\bibfield  {journal} {\bibinfo  {journal}
  {Adv. At. Mol. Opt. Phys.}\ }\textbf {\bibinfo {volume} {60}},\ \bibinfo
  {pages} {315} (\bibinfo {year} {2011})}\BibitemShut {NoStop}%
\bibitem [{\citenamefont {Izrailev}(1990)}]{Izr1990}%
  \BibitemOpen
  \bibfield  {author} {\bibinfo {author} {\bibfnamefont {F.~M.}\ \bibnamefont
  {Izrailev}},\ }\href {\doibase DOI: 10.1016/0370-1573(90)90067-C} {\bibfield
  {journal} {\bibinfo  {journal} {Physics Reports}\ }\textbf {\bibinfo {volume}
  {196}},\ \bibinfo {pages} {299 } (\bibinfo {year} {1990})}\BibitemShut
  {NoStop}%
  \bibitem{nonl}
  S. Wimberger, I. Guarneri, and S. Fishman, Nonlinearity {\bf 16}, 1381 (2003).
  
  \bibitem [{\citenamefont {Deng}\ \emph {et~al.}(1999)\citenamefont {Deng},
  \citenamefont {Hagley}, \citenamefont {Denschlag}, \citenamefont {Simsarian},
  \citenamefont {Edwards}, \citenamefont {Clark}, \citenamefont {Helmerson},
  \citenamefont {Rolston},\ and\ \citenamefont
  {Phillips}}]{PhysRevLett.83.5407}%
  \BibitemOpen
  \bibfield  {author} {\bibinfo {author} {\bibfnamefont {L.}~\bibnamefont
  {Deng}}, \bibinfo {author} {\bibfnamefont {E.~W.}\ \bibnamefont {Hagley}},
  \bibinfo {author} {\bibfnamefont {J.}~\bibnamefont {Denschlag}}, \bibinfo
  {author} {\bibfnamefont {J.~E.}\ \bibnamefont {Simsarian}}, \bibinfo {author}
  {\bibfnamefont {M.}~\bibnamefont {Edwards}}, \bibinfo {author} {\bibfnamefont
  {C.~W.}\ \bibnamefont {Clark}}, \bibinfo {author} {\bibfnamefont
  {K.}~\bibnamefont {Helmerson}}, \bibinfo {author} {\bibfnamefont {S.~L.}\
  \bibnamefont {Rolston}}, \ and\ \bibinfo {author} {\bibfnamefont {W.~D.}\
  \bibnamefont {Phillips}},\ }\href {\doibase 10.1103/PhysRevLett.83.5407}
  {\bibfield  {journal} {\bibinfo  {journal} {Phys. Rev. Lett.}\ }\textbf
  {\bibinfo {volume} {83}},\ \bibinfo {pages} {5407} (\bibinfo {year}
  {1999})}\BibitemShut {NoStop}%
\bibitem [{\citenamefont {Sadgrove}\ \emph {et~al.}(2007)\citenamefont
  {Sadgrove}, \citenamefont {Horikoshi}, \citenamefont {Sekimura},\ and\
  \citenamefont {Nakagawa}}]{sadgrove2007}%
  \BibitemOpen
  \bibfield  {author} {\bibinfo {author} {\bibfnamefont {M.}~\bibnamefont
  {Sadgrove}}, \bibinfo {author} {\bibfnamefont {M.}~\bibnamefont {Horikoshi}},
  \bibinfo {author} {\bibfnamefont {T.}~\bibnamefont {Sekimura}}, \ and\
  \bibinfo {author} {\bibfnamefont {K.}~\bibnamefont {Nakagawa}},\ }\href
  {\doibase 10.1103/PhysRevLett.99.043002} {\bibfield  {journal} {\bibinfo
  {journal} {Phys. Rev. Lett.}\ }\textbf {\bibinfo {volume} {99}},\ \bibinfo
  {pages} {043002} (\bibinfo {year} {2007})}\BibitemShut {NoStop}%
\bibitem [{\citenamefont {Dana}\ \emph {et~al.}(2008)\citenamefont {Dana},
  \citenamefont {Ramareddy}, \citenamefont {Talukdar},\ and\ \citenamefont
  {Summy}}]{gil2008}%
  \BibitemOpen
  \bibfield  {author} {\bibinfo {author} {\bibfnamefont {I.}~\bibnamefont
  {Dana}}, \bibinfo {author} {\bibfnamefont {V.}~\bibnamefont {Ramareddy}},
  \bibinfo {author} {\bibfnamefont {I.}~\bibnamefont {Talukdar}}, \ and\
  \bibinfo {author} {\bibfnamefont {G.~S.}\ \bibnamefont {Summy}},\ }\href
  {\doibase 10.1103/PhysRevLett.100.024103} {\bibfield  {journal} {\bibinfo
  {journal} {Phys. Rev. Lett.}\ }\textbf {\bibinfo {volume} {100}},\ \bibinfo
  {pages} {024103} (\bibinfo {year} {2008})}\BibitemShut {NoStop}%
\bibitem [{\citenamefont {Shrestha}\ \emph {et~al.}(2012)\citenamefont
  {Shrestha}, \citenamefont {Ni}, \citenamefont {Lam}, \citenamefont
  {Wimberger},\ and\ \citenamefont {Summy}}]{gil2012}%
  \BibitemOpen
  \bibfield  {author} {\bibinfo {author} {\bibfnamefont {R.~K.}\ \bibnamefont
  {Shrestha}}, \bibinfo {author} {\bibfnamefont {J.}~\bibnamefont {Ni}},
  \bibinfo {author} {\bibfnamefont {W.~K.}\ \bibnamefont {Lam}}, \bibinfo
  {author} {\bibfnamefont {S.}~\bibnamefont {Wimberger}}, \ and\ \bibinfo
  {author} {\bibfnamefont {G.~S.}\ \bibnamefont {Summy}},\ }\href {\doibase
  10.1103/PhysRevA.86.043617} {\bibfield  {journal} {\bibinfo  {journal} {Phys.
  Rev. A}\ }\textbf {\bibinfo {volume} {86}},\ \bibinfo {pages} {043617}
  (\bibinfo {year} {2012})}\BibitemShut {NoStop}%
\bibitem [{\citenamefont {Sadgrove}\ and\ \citenamefont
  {Wimberger}(2009)}]{SadgroveWimberger2009}%
  \BibitemOpen
  \bibfield  {author} {\bibinfo {author} {\bibfnamefont {M.}~\bibnamefont
  {Sadgrove}}\ and\ \bibinfo {author} {\bibfnamefont {S.}~\bibnamefont
  {Wimberger}},\ }\href {http://stacks.iop.org/1367-2630/11/i=8/a=083027}
  {\bibfield  {journal} {\bibinfo  {journal} {New Journal of Physics}\ }\textbf
  {\bibinfo {volume} {11}},\ \bibinfo {pages} {083027} (\bibinfo {year}
  {2009})}\BibitemShut {NoStop}%
\bibitem{BK2004}
O. Buerschaper and K. Burnett, quant-ph/0406039 (unpublished).

\bibitem{QAM2006}
Z.-Y. Ma, K. Burnett, M. B. d'Arcy, and A. S. Gardiner, 
Phys. Rev. A {\bf 73}, 013401 (2006).

\bibitem [{\citenamefont {Sadgrove}\ \emph {et~al.}(2005)\citenamefont
  {Sadgrove}, \citenamefont {Wimberger}, \citenamefont {Parkins},\ and\
  \citenamefont {Leonhardt}}]{Sadgrove_2005}%
  \BibitemOpen
  \bibfield  {author} {\bibinfo {author} {\bibfnamefont {M.}~\bibnamefont
  {Sadgrove}}, \bibinfo {author} {\bibfnamefont {S.}~\bibnamefont {Wimberger}},
  \bibinfo {author} {\bibfnamefont {S.}~\bibnamefont {Parkins}}, \ and\
  \bibinfo {author} {\bibfnamefont {R.}~\bibnamefont {Leonhardt}},\ }\href
  {\doibase 10.1103/physrevlett.94.174103} {\bibfield  {journal} {\bibinfo
  {journal} {Phys. Rev. Lett.}\ }\textbf {\bibinfo {volume} {94}},\ \bibinfo
  {pages} {174103} (\bibinfo {year} {2005})}\BibitemShut {NoStop}%
\bibitem [{\citenamefont {Behinaein}\ \emph {et~al.}(2006)\citenamefont
  {Behinaein}, \citenamefont {Ramareddy}, \citenamefont {Ahmadi},\ and\
  \citenamefont {Summy}}]{Behinaein_2006}%
  \BibitemOpen
  \bibfield  {author} {\bibinfo {author} {\bibfnamefont {G.}~\bibnamefont
  {Behinaein}}, \bibinfo {author} {\bibfnamefont {V.}~\bibnamefont
  {Ramareddy}}, \bibinfo {author} {\bibfnamefont {P.}~\bibnamefont {Ahmadi}}, \
  and\ \bibinfo {author} {\bibfnamefont {G.~S.}\ \bibnamefont {Summy}},\ }\href
  {\doibase 10.1103/physrevlett.97.244101} {\bibfield  {journal} {\bibinfo
  {journal} {Phys. Rev. Lett.}\ }\textbf {\bibinfo {volume} {97}},\ \bibinfo
  {pages} {244101} (\bibinfo {year} {2006})}\BibitemShut {NoStop}%
\bibitem [{\citenamefont {Talukdar}\ \emph {et~al.}(2010)\citenamefont
  {Talukdar}, \citenamefont {Shrestha},\ and\ \citenamefont
  {Summy}}]{Talukdar_2010}%
  \BibitemOpen
  \bibfield  {author} {\bibinfo {author} {\bibfnamefont {I.}~\bibnamefont
  {Talukdar}}, \bibinfo {author} {\bibfnamefont {R.}~\bibnamefont {Shrestha}},
  \ and\ \bibinfo {author} {\bibfnamefont {G.~S.}\ \bibnamefont {Summy}},\
  }\href {\doibase 10.1103/physrevlett.105.054103} {\bibfield  {journal}
  {\bibinfo  {journal} {Phys. Rev. Lett.}\ }\textbf {\bibinfo {volume} {105}},\
  \bibinfo {pages} {054103} (\bibinfo {year} {2010})}\BibitemShut {NoStop}%
\bibitem [{\citenamefont {Shrestha}\ \emph {et~al.}(2013)\citenamefont
  {Shrestha}, \citenamefont {Wimberger}, \citenamefont {Ni}, \citenamefont
  {Lam},\ and\ \citenamefont {Summy}}]{gil2013a}%
  \BibitemOpen
  \bibfield  {author} {\bibinfo {author} {\bibfnamefont {R.~K.}\ \bibnamefont
  {Shrestha}}, \bibinfo {author} {\bibfnamefont {S.}~\bibnamefont {Wimberger}},
  \bibinfo {author} {\bibfnamefont {J.}~\bibnamefont {Ni}}, \bibinfo {author}
  {\bibfnamefont {W.~K.}\ \bibnamefont {Lam}}, \ and\ \bibinfo {author}
  {\bibfnamefont {G.~S.}\ \bibnamefont {Summy}},\ }\href {\doibase
  10.1103/PhysRevE.87.020902} {\bibfield  {journal} {\bibinfo  {journal} {Phys.
  Rev. E}\ }\textbf {\bibinfo {volume} {87}},\ \bibinfo {pages} {020902}
  (\bibinfo {year} {2013})}\BibitemShut {NoStop}%
\bibitem [{\citenamefont {Ryu}\ \emph {et~al.}(2006)\citenamefont {Ryu},
  \citenamefont {Andersen}, \citenamefont {Vaziri}, \citenamefont {d'Arcy},
  \citenamefont {Grossman}, \citenamefont {Helmerson},\ and\ \citenamefont
  {Phillips}}]{Ryu}%
  \BibitemOpen
  \bibfield  {author} {\bibinfo {author} {\bibfnamefont {C.}~\bibnamefont
  {Ryu}}, \bibinfo {author} {\bibfnamefont {M.~F.}\ \bibnamefont {Andersen}},
  \bibinfo {author} {\bibfnamefont {A.}~\bibnamefont {Vaziri}}, \bibinfo
  {author} {\bibfnamefont {M.~B.}\ \bibnamefont {d'Arcy}}, \bibinfo {author}
  {\bibfnamefont {J.~M.}\ \bibnamefont {Grossman}}, \bibinfo {author}
  {\bibfnamefont {K.}~\bibnamefont {Helmerson}}, \ and\ \bibinfo {author}
  {\bibfnamefont {W.~D.}\ \bibnamefont {Phillips}},\ }\href {\doibase
  10.1103/PhysRevLett.96.160403} {\bibfield  {journal} {\bibinfo  {journal}
  {Phys. Rev. Lett.}\ }\textbf {\bibinfo {volume} {96}},\ \bibinfo {pages}
  {160403} (\bibinfo {year} {2006})}\BibitemShut {NoStop}%
  \bibitem{gil2013b}
  R. K. Shrestha, J. Ni, W. K. Lam, G. S. Summy, and S. Wimberger, 
  Phys. Rev. E {\bf 88}, 034901 (2013).
  
\bibitem [{\citenamefont {Schmitz}\ \emph {et~al.}(2009)\citenamefont
  {Schmitz}, \citenamefont {Matjeschk}, \citenamefont {Schneider},
  \citenamefont {Glueckert}, \citenamefont {Enderlein}, \citenamefont {Huber},\
  and\ \citenamefont {Sch\"atz}}]{PhysRevLett.103.090504}%
  \BibitemOpen
  \bibfield  {author} {\bibinfo {author} {\bibfnamefont {H.}~\bibnamefont
  {Schmitz}}, \bibinfo {author} {\bibfnamefont {R.}~\bibnamefont {Matjeschk}},
  \bibinfo {author} {\bibfnamefont {C.}~\bibnamefont {Schneider}}, \bibinfo
  {author} {\bibfnamefont {J.}~\bibnamefont {Glueckert}}, \bibinfo {author}
  {\bibfnamefont {M.}~\bibnamefont {Enderlein}}, \bibinfo {author}
  {\bibfnamefont {T.}~\bibnamefont {Huber}}, \ and\ \bibinfo {author}
  {\bibfnamefont {T.}~\bibnamefont {Sch\"atz}},\ }\href {\doibase
  10.1103/PhysRevLett.103.090504} {\bibfield  {journal} {\bibinfo  {journal}
  {Phys. Rev. Lett.}\ }\textbf {\bibinfo {volume} {103}},\ \bibinfo {pages}
  {090504} (\bibinfo {year} {2009})}\BibitemShut {NoStop}%
\bibitem [{\citenamefont {Broome}\ \emph {et~al.}(2010)\citenamefont {Broome},
  \citenamefont {Fedrizzi}, \citenamefont {Lanyon}, \citenamefont {Kassal},
  \citenamefont {Aspuru-Guzik},\ and\ \citenamefont
  {White}}]{PhysRevLett.104.153602}%
  \BibitemOpen
  \bibfield  {author} {\bibinfo {author} {\bibfnamefont {M.~A.}\ \bibnamefont
  {Broome}}, \bibinfo {author} {\bibfnamefont {A.}~\bibnamefont {Fedrizzi}},
  \bibinfo {author} {\bibfnamefont {B.~P.}\ \bibnamefont {Lanyon}}, \bibinfo
  {author} {\bibfnamefont {I.}~\bibnamefont {Kassal}}, \bibinfo {author}
  {\bibfnamefont {A.}~\bibnamefont {Aspuru-Guzik}}, \ and\ \bibinfo {author}
  {\bibfnamefont {A.~G.}\ \bibnamefont {White}},\ }\href {\doibase
  10.1103/PhysRevLett.104.153602} {\bibfield  {journal} {\bibinfo  {journal}
  {Phys. Rev. Lett.}\ }\textbf {\bibinfo {volume} {104}},\ \bibinfo {pages}
  {153602} (\bibinfo {year} {2010})}\BibitemShut {NoStop}%
\bibitem [{\citenamefont {Z\"ahringer}\ \emph {et~al.}(2010)\citenamefont
  {Z\"ahringer}, \citenamefont {Kirchmair}, \citenamefont {Gerritsma},
  \citenamefont {Solano}, \citenamefont {Blatt},\ and\ \citenamefont
  {Roos}}]{PhysRevLett.104.100503}%
  \BibitemOpen
  \bibfield  {author} {\bibinfo {author} {\bibfnamefont {F.}~\bibnamefont
  {Z\"ahringer}}, \bibinfo {author} {\bibfnamefont {G.}~\bibnamefont
  {Kirchmair}}, \bibinfo {author} {\bibfnamefont {R.}~\bibnamefont
  {Gerritsma}}, \bibinfo {author} {\bibfnamefont {E.}~\bibnamefont {Solano}},
  \bibinfo {author} {\bibfnamefont {R.}~\bibnamefont {Blatt}}, \ and\ \bibinfo
  {author} {\bibfnamefont {C.~F.}\ \bibnamefont {Roos}},\ }\href {\doibase
  10.1103/PhysRevLett.104.100503} {\bibfield  {journal} {\bibinfo  {journal}
  {Phys. Rev. Lett.}\ }\textbf {\bibinfo {volume} {104}},\ \bibinfo {pages}
  {100503} (\bibinfo {year} {2010})}\BibitemShut {NoStop}%
\bibitem [{\citenamefont {Sansoni}\ \emph {et~al.}(2012)\citenamefont
  {Sansoni}, \citenamefont {Sciarrino}, \citenamefont {Vallone}, \citenamefont
  {Mataloni}, \citenamefont {Crespi}, \citenamefont {Ramponi},\ and\
  \citenamefont {Osellame}}]{PhysRevLett.108.010502}%
  \BibitemOpen
  \bibfield  {author} {\bibinfo {author} {\bibfnamefont {L.}~\bibnamefont
  {Sansoni}}, \bibinfo {author} {\bibfnamefont {F.}~\bibnamefont {Sciarrino}},
  \bibinfo {author} {\bibfnamefont {G.}~\bibnamefont {Vallone}}, \bibinfo
  {author} {\bibfnamefont {P.}~\bibnamefont {Mataloni}}, \bibinfo {author}
  {\bibfnamefont {A.}~\bibnamefont {Crespi}}, \bibinfo {author} {\bibfnamefont
  {R.}~\bibnamefont {Ramponi}}, \ and\ \bibinfo {author} {\bibfnamefont
  {R.}~\bibnamefont {Osellame}},\ }\href {\doibase
  10.1103/PhysRevLett.108.010502} {\bibfield  {journal} {\bibinfo  {journal}
  {Phys. Rev. Lett.}\ }\textbf {\bibinfo {volume} {108}},\ \bibinfo {pages}
  {010502} (\bibinfo {year} {2012})}\BibitemShut {NoStop}%
\bibitem [{\citenamefont {Alberti}\ \emph {et~al.}(2014)\citenamefont
  {Alberti}, \citenamefont {Alt}, \citenamefont {Werner},\ and\ \citenamefont
  {Meschede}}]{njpBonn}%
  \BibitemOpen
  \bibfield  {author} {\bibinfo {author} {\bibfnamefont {A.}~\bibnamefont
  {Alberti}}, \bibinfo {author} {\bibfnamefont {W.}~\bibnamefont {Alt}},
  \bibinfo {author} {\bibfnamefont {R.}~\bibnamefont {Werner}}, \ and\ \bibinfo
  {author} {\bibfnamefont {D.}~\bibnamefont {Meschede}},\ }\href@noop {}
  {\bibfield  {journal} {\bibinfo  {journal} {New Journal of Physics}\ }\textbf
  {\bibinfo {volume} {16}},\ \bibinfo {pages} {123052} (\bibinfo {year}
  {2014})}\BibitemShut {NoStop}%
\bibitem [{\citenamefont {Ra}\ \emph {et~al.}(2013)\citenamefont {Ra},
  \citenamefont {Tichy}, \citenamefont {Lim}, \citenamefont {Kwon},
  \citenamefont {Mintert}, \citenamefont {Buchleitner},\ and\ \citenamefont
  {Kim}}]{Ra22012013}%
  \BibitemOpen
  \bibfield  {author} {\bibinfo {author} {\bibfnamefont {Y.-S.}\ \bibnamefont
  {Ra}}, \bibinfo {author} {\bibfnamefont {M.~C.}\ \bibnamefont {Tichy}},
  \bibinfo {author} {\bibfnamefont {H.-T.}\ \bibnamefont {Lim}}, \bibinfo
  {author} {\bibfnamefont {O.}~\bibnamefont {Kwon}}, \bibinfo {author}
  {\bibfnamefont {F.}~\bibnamefont {Mintert}}, \bibinfo {author} {\bibfnamefont
  {A.}~\bibnamefont {Buchleitner}}, \ and\ \bibinfo {author} {\bibfnamefont
  {Y.-H.}\ \bibnamefont {Kim}},\ }\href {\doibase 10.1073/pnas.1206910110}
  {\bibfield  {journal} {\bibinfo  {journal} {Proceedings of the National
  Academy of Sciences}\ }\textbf {\bibinfo {volume} {110}},\ \bibinfo {pages}
  {1227} (\bibinfo {year} {2013})}\BibitemShut {NoStop}%
\bibitem [{\citenamefont {Spagnolo}\ and\ \citenamefont {{\em et
  al.}}(2014)}]{bosonsample2}%
  \BibitemOpen
  \bibfield  {author} {\bibinfo {author} {\bibfnamefont {N.}~\bibnamefont
  {Spagnolo}}\ and\ \bibinfo {author} {\bibnamefont {{\em et al.}}},\
  }\href@noop {} {\bibfield  {journal} {\bibinfo  {journal} {Nature Photonics}\
  }\textbf {\bibinfo {volume} {8}},\ \bibinfo {pages} {615} (\bibinfo {year}
  {2014})}\BibitemShut {NoStop}%
\bibitem [{\citenamefont {Tichy}\ \emph {et~al.}(2014)\citenamefont {Tichy},
  \citenamefont {Mayer}, \citenamefont {Buchleitner},\ and\ \citenamefont
  {M\o{}lmer}}]{bosonsample5}%
  \BibitemOpen
  \bibfield  {author} {\bibinfo {author} {\bibfnamefont {M.~C.}\ \bibnamefont
  {Tichy}}, \bibinfo {author} {\bibfnamefont {K.}~\bibnamefont {Mayer}},
  \bibinfo {author} {\bibfnamefont {A.}~\bibnamefont {Buchleitner}}, \ and\
  \bibinfo {author} {\bibfnamefont {K.}~\bibnamefont {M\o{}lmer}},\ }\href
  {\doibase 10.1103/PhysRevLett.113.020502} {\bibfield  {journal} {\bibinfo
  {journal} {Phys. Rev. Lett.}\ }\textbf {\bibinfo {volume} {113}},\ \bibinfo
  {pages} {020502} (\bibinfo {year} {2014})}\BibitemShut {NoStop}%
\end{thebibliography}

\end{document}